\documentclass{article}

\usepackage{graphicx}
\usepackage{amssymb, amsmath, amsthm}   
\usepackage[round]{natbib}
\usepackage{setspace}

\newcommand{\bE}{{\mathbb E}}

\newcommand{\bP}{{\mathbb P}}

\newcommand{\shuff}{s}
\newcommand{\Pheidole}{\emph{Pheidole}}

\newcommand{\intNode}{N}
\newcommand{\run}{{\mathcal R}}

\newcommand{\eat}[1]{}
\newcommand{\eatfig}[1]{#1}
\newcommand{\eatend}[1]{}

\newcommand{\SBsection}[1]{\vspace{.2cm} \noindent \begin{center} \textsc{#1} \end{center} \vspace{-.05cm}}
\newcommand{\SBsubsection}[1]{\vspace{.0cm} \noindent \begin{center} \textit{#1} \end{center} \vspace{-.1cm}}
\newcommand{\SBsubsubsection}[1]{\vspace{.4cm} \noindent \begin{center} \textsc{\small #1} \end{center} \vspace{0cm}}

\newtheorem{thm}{Theorem}
\newtheorem{obs}[thm]{Observation}
\newtheorem{lem}[thm]{Lemma}

\newtheorem{cor}[thm]{Corollary}
\newtheorem{prop}[thm]{Proposition}

\theoremstyle{definition}
\newtheorem{defi}[thm]{Definition}
\theoremstyle{definition}

\theoremstyle{definition}

\begin{document}

\begin{center}

\emph{Running head:} 

INVESTIGATING RELATIVE TIMING ON PHYLOGENETIC TREES

\vspace{.5cm}

\LARGE{A method for investigating relative timing information on phylogenetic trees}

\vspace{0.75cm}
\large{

Daniel Ford$^{\ast}$\\
\textsl{
Google Inc.\\
1600 Amphitheatre Parkway\\
Mountain View, CA 94043\\
USA
}

\vspace{0.5cm}
Tanja Gernhard$^{\ast}$\\
\textsl{
Kombinatorische Geometrie (M9)\\
Zentrum Mathematik,\\
Technische Universit\"{a}t M\"{u}nchen\\
Boltzmannstr. 3,\\
85747 Garching bei M\"{u}nchen\\
Germany
}

\vspace{0.5cm}
Frederick A. Matsen$^{\ast}$\\
\textsl{
Department of Statistics\\
University of California, Berkeley\\
367 Evans Hall \#429\\
Berkeley, CA 94720-3860\\
USA\\
http://www.stat.berkeley.edu/$\tilde{\ }$matsen/ 
}

\vspace{0.5cm}
$^{\ast}$All authors contributed equally to this manuscript.
}

\vspace{0.5cm}
\textsl{
Corresponding Author:\\
Frederick A. Matsen\\
phone: +1 510 642 2450 \\
fax: +1 510 642 7892 \\
email: matsen@berkeley.edu\\
}

\end{center}

\vspace{0.5cm} 


\newpage

\begin{spacing}{1}

\begin{abstract}
  In this paper we present a new way to understand the timing of
  branching events in phylogenetic trees. Our method explicitly
  considers the relative timing of diversification events between
  sister clades; as such it is complimentary to existing methods using
  lineages-through-time plots which consider diversification in
  aggregate. The method looks for evidence of diversification
  happening in lineage-specific ``bursts'', or the opposite, where
  diversification between two clades happens in an unusually regular
  fashion. In order to be able to distinguish interesting events from
  stochasticity, we propose two classes of neutral models on trees
  with timing information and develop a statistical framework for
  testing these models. Our models substantially generalize both the
  coalescent with ancestral population size variation and the
  global-rate speciation-extinction models. We end the paper with
  several
  example applications: first, we show that the evolution of the
  Hepatitis C virus appears to proceed in a lineage-specific bursting
  fashion. Second, we analyze a large tree of ants, demonstrating that
  a period of elevated diversification rates does not appear to
  occurred in a bursting manner.
\end{abstract}

\SBsection{Introduction}
\label{sec:intro}

Understanding the tempo and mode of diversification is one of the major challenges of
evolutionary biology. Phylogenetic trees with timing information are
powerful tools for answering questions about tempo and mode. 
Such trees were once available only in situations with a rich fossil
record, where the timing information might have come from
radiocarbon dating or stratigraphic information.
However, modern techniques of phylogenetic analysis are capable of
reconstructing not only the topology of phylogenetic trees, but can
also reconstruct information about the timing of diversification events even when
limited or no fossil evidence is available. This can be done in one of
a number of ways. One can first test if a molecular clock is
appropriate [see \citet{felsensteinBook} p. 323], then reconstruct
under the assumption of a molecular clock. One can reconstruct a tree
with branch lengths using any method and then apply rate smoothing
\citep{sandersonR8s03}. One may also choose from the variety of
``relaxed clock'' methods which allow the rate of substitution to vary
within the tree \citep{gillespieEpisodicClock84,
huelsenbeckEaRelaxing00, drummondEaRelaxed06}. Of course, the accuracy
of any these techniques depend on a correct choice of model and a
strong phylogenetic signal along with perhaps some fossil calibration
points.

Phylogenetic trees with timing information can then be used to make inferences about
the forces guiding the evolution of the taxa. For example, the
paper of \citet{moreauEaAnts06} notes that there was
a period of high diversification rate in ant lineages during the rise of
angiosperms.  Another paper by \citet{harmonEaTempoMode03}
uses the deviation of four groups of lizards from the pure-birth
model of diversification to make inferences about their evolutionary
radiations. 

Given the number of methods available for reconstructing phylogenetic
trees with diversification timing information, and the interest in
investigating temporal properties of those trees, the number of direct methods
available to investigate timing information on phylogenetic trees is surprisingly small.
The most popular ways of investigating timing in phylogenetic trees
are lineage-through-time (LTT) plots and the associated $\gamma$
statistic introduced by \citet{incompletePybusHarvey00} [for a helpful
review article see \citep{trendsRicklefs07}]. LTT plots
have time $t$ on the $x$ axis and simply show the number of lineages
which were present in the phylogenetic tree at time $t$ on the $y$
axis. A constant rate pure-birth process would have the number of taxa
increasing exponentially; it is common to compare LTT plots to an
exponential curve \citep{zinkSlowinskiAvian95, harmonEaTempoMode03}.
The $\gamma$ statistic is computed based on the periods during which the LTT plot
stays constant (called the ``inter-node intervals''); the $\gamma$ for
a pure-birth diversification process will have a standard normal distribution. Broadly
speaking $\gamma < 0$ implies that diversification rates were high
early in history, while $\gamma > 0$ implies that most diversification
has happened more recently. A similar statistic with the same goals in
mind was constructed by \citet{zinkSlowinskiAvian95}. 

However, much more information is available in a phylogenetic tree
with diversification timing information than can be summarized in a
LTT plot or a derivative statistic. Consider the tree in
Figure~1, with two sets of sister taxa, $A$
and $B$. The taxa in $B$ had a period of relatively high
diversification rate early in evolutionary history, during which time the lineage
leading to $A$ is in a period of stasis. Then lineage $A$ experiences
a burst of diversification, and the taxa in $B$ do not experience any
lineage-splitting events during this time. 
We will call the sort of diversification seen in
Figure~1 ``lineage-specific bursting'' (LSB) diversification, i.e. where the
relative diversification rates in two sister clades vary over time.

\eatfig{
\begin{figure}
  \begin{center}
    \includegraphics[width=4cm]{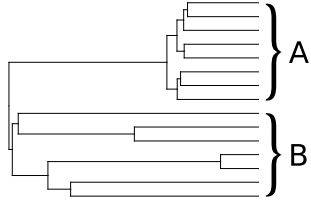}
  \end{center}
  \caption{A motivating example showing ``bursting'' diversification.  Namely, in
the oldest part of the tree, diversification events happen exclusively
in the $B$ lineage, followed by a period of high diversification rate
in the $A$ lineage. This paper constructs a statistical framework for
analyzing such ``bursting'' patterns or their opposite.}
  \label{fig:motiv}
\end{figure}
}

The lineage-specific bursting diversification seen in
Figure~1 would not be apparent in an LTT plot.  Indeed,
LTT plots take the timing information out of the context of the
phylogenetic tree from which from which they are derived, and thus ignore
information about how the timings relate to topology of the tree. This
context can be crucial, as we now argue. 

One would like to be able to say if, for example, the pattern seen in
Figure~1 arose simply ``by chance.'' In order to do so, we
need two things: first, a convenient way to summarize the timing
information, and second, a set of null models which define what we
mean with ``by chance.'' For a given internal node, we summarize the
timing information at that node by writing down the order of diversification events
by clade. For instance, we associate with the root node of
Figure~1 the sequence
$\shuff = BBBBBAAAAAAAB$ which we will call a ``shuffle'' in analogy to a
shuffling of cards labeled $A$ and $B$. We make a more formal
definition of shuffles in the section labeled ``Tree Shuffles.''

Now that we have summarized the timing information at the root node as
a shuffle $\shuff$,
we would like to think about if $\shuff$ arose
``by chance.'' This of course requires us to define a probability
distribution on shuffles; we demonstrate below that a wide class of
null models on phylogenetic trees give the uniform distribution on
shuffles. The uniform distribution in this setting is what one would
get by throwing the $A$'s and $B$'s of the shuffle into a bag and
drawing them out one by one uniformly. Thus it seems reasonably
unlikely that the shuffle $\shuff$ would arise by chance, having first
a long run of $B$'s then a long run of $A$'s.

We can attach a $p$-value to a shuffle by using the
``runs distribution.'' The number of ``runs'' is simply the number of
sequences of the same letter: in this case, there is a run of $B$'s,
then a run of $A$'s, then another $B$. That totals three runs. Under
the uniform distribution, the
probability of seeing a given number of runs in this setting is
known from classical statistics, and can be calculated via
Equation (\ref{EqnRunDistr}). The probability of seeing 3 runs with 6 $A$'s
and 7 $B$'s is about 0.00641, and the probability of seeing 2 runs is
about 0.00117. We can interpret the sum of these two probabilities,
0.00758, as the significance level of the LSB diversification seen in
Figure~1. Being below the 1\% significance level, we can
interpret this shuffle as being quite significant; thus if the tree in
Figure~1 came from data, the observed lineage-specific
diversification might require some explanation. Please note that for
simplicity this example only considers the root shuffle; however the
main body of the paper is dedicated to investigating all shuffles
simultaneously.

The first aim of this paper is to provide analytical tools to compare
diversification rates between lineages. In doing so, we hope to
provide a complimentary perspective to that provided by LTT plots and
associated statistics. 
In particular, our method can detect
``lineage-specific bursting'' (LSB) diversification, i.e. where the
diversification rates in two sister clades vary over time. One might
expect LSB diversification if a lineage diversifies to fill variants
of a single niche, or if a key innovation appears which makes
further diversifications more likely. By comparing
the results of our analysis to results using LLT plots, we may be able
to tease apart causes of diversification rate changes--- are they
lineage-specific or due to global events?

The second aim of this paper is to investigate null models of phylogenetic
trees with timing information. In contrast to the setting of
phylogenetic tree shape, where a number of models are available 
\citep{aldousCladograms95, fordAlpha05, mooersEaTreeShapeModels07},
there are relatively few models available for trees with timing
information. Null models are important as they allow us to distinguish between
stochastic sampling and actual events which need investigation; they
are thus important tools for assessing significance.

We conclude the paper with example applications. Our first
example application uses Hepatitis C (HCV) data, and shows that trees from
this data demonstrate a limited but significant amount of LSB
diversification. This analysis may imply a note of caution for
researchers using coalescent methods to analyze HCV data. Our second
application is to the ant data of \citep{moreauEaAnts06} and
\citep{moreauPheidole},
the lineages of which do not appear to demonstrate significant LSB diversification,
despite some other interesting characteristics of their history.

Our paper is one contribution to the area of understanding mechanisms
of diversification from phylogenetic trees. Besides lineages through
time plots and $\gamma$, there is an entire
literature on phylogenetic tree shape (which does not include branch
length); for an excellent review see \citet{mooersHeard97}. 
There are also a number of interesting papers which use trait
information, for example 
\citet{pagelInferring97} and \citet{reeDetecting05}. However, our method
is the first to use just a phylogenetic tree with branch lengths in a
way which integrates both sources of information.

\eat{
Before launching into the manuscript proper, it may be useful to
consider the meaning of the word ``neutral.'' There are two senses in
which this word can be used. First there is the wide sense, where it
is defined as ``conforming to some neutral model.'' However, in the
population genetics literature it has come to have a more narrow
meaning i.e. ``conforming to the constant population size
coalescent.'' For example, Tajima's D statistic is a test of
neutrality in the narrow sense.
}

\SBsection{Tree shuffles}
\label{sec:ranked}

Our method is based on ``ranked'' phylogenetic trees: trees for which
the order of branching events in the tree is specified in a way
compatible with the topology (more specific definition
below). Such trees have
been called ``dendrograms'' \citep{page91}. As we discuss
below, a ranked phylogenetic tree is equivalent to a
phylogenetic tree with a ``shuffle'' at each internal node
specifying relative timing information.  As described below,
a broad class of
neutral diversification models give the uniform distribution on
shuffles, which leads to some natural tests for deviation from
these models. Thus evidence of deviation from the uniform distribution
on shuffles is evidence of deviation from this entire broad class of
neutral models. (Note that by ``model'' we mean a forward time ranked-tree-valued
 stochastic process.  Often, a model with branch lengths
is given, in which case we consider the induced model given by
considering ranks.)

The intuition behind the shuffle idea is presented in
Figure~2. As shown in this figure, the relative order
of bifurcation events for an internal node of a tree is determined by
the sequence of full and hollow circles on the left side of each tree.
We call this sequence a ``shuffle.''  Shuffles also have a natural
interpretation in terms of evolutionary history. Namely, ``bursting''
diversification leads to symbols of a shuffle clustering together. The
opposite situation, where there is a post-diversification delay before
a lineage can diversify again, can be recognized by the interspersing
of different symbols. This latter situation has been called
``refractory'' diversification \citep{stumpedLososAdler95}.

\eatfig{
\begin{figure}
  \begin{center}
    \includegraphics[width=12cm]{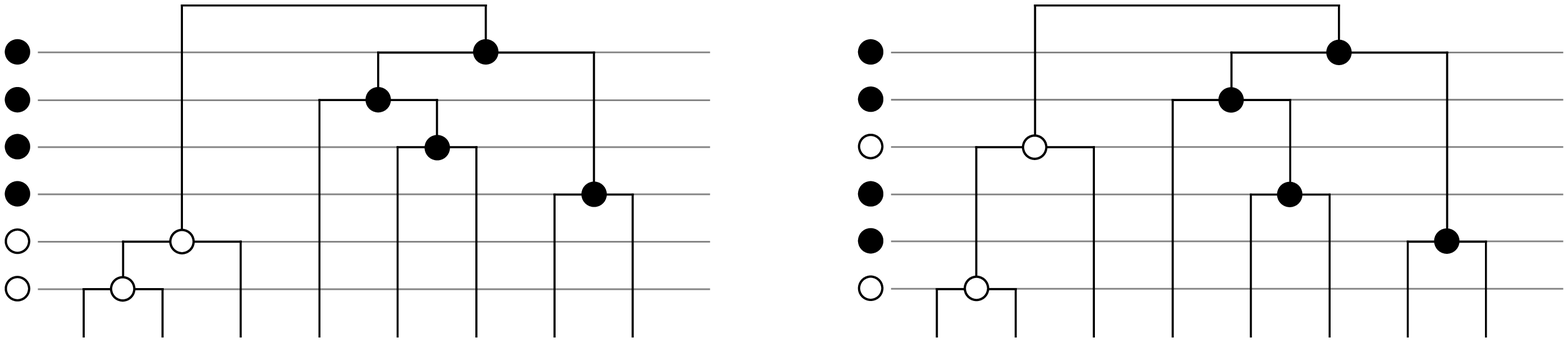}
  \end{center}
  \caption{A shuffle at a given internal
  node. Bifurcations on the left subtree are marked with a hollow
  circle, and those on the right subtree are marked with a solid
  circle. The relative timing for these events is shown beside the
  tree; we call this sequence of symbols a ``shuffle.'' A set of
  shuffles for every internal node of a phylogenetic tree
  exactly determines the relative order of bifurcation events. Similar
  type symbols occurring together as in the left tree is evidence of
  lineage-specific bursts.}
  \label{fig:shuffle}
\end{figure}
}

We now make more formal definitions of our terms.
For the purposes of this paper, a {\em phylogenetic tree} is a rooted tree with distinct leaf
labels. We will denote the set of interior nodes of a phylogenetic
tree $T$ with $\intNode_T$. 
For an internal node $v$ in $\intNode_T$, define $T_v$ to be 
the rooted subtree of $T$ containing all the descendants of $v$. The
``daughter trees'' of $v$ are the two subtrees of $T_v$
which we obtain by deleting $v$ and its two incident edges.
For the first part of the paper, we assume that the phylogenetic trees
are bifurcating. We describe later how to
generalize the ideas presented to the case of multifurcating trees.
A {\em rank function} on an arbitrary set $S$ is simply an ordering of
the elements of that set; mathematically it is a one-to-one
mapping from $S$ to ranks $\{1,2,\ldots,|S|\}$. 
A {\em rank function on a phylogenetic tree} $T$ is a rank
function on the set of interior vertices $\intNode_T$ with
the property that the ranks are increasing on any path from the root
to a leaf. We call a phylogenetic tree with a rank function a \emph{ranked
phylogenetic tree} or simply \emph{ranked tree} \citep{Steel2003}.

In mathematics, a {\em total order} on a set is simply a binary relation
(usually written $<$) such that for any two distinct elements $a$ and
$b$ of the set either $a < b$ or $b < a$.  Note that
a rank function on a set is equivalent to a total order on that set:
given a total order one can rank the elements in increasing order of
rank, and given a rank function one can define a total order by
numerical inequality of rank. Thus a
ranked phylogenetic tree is exactly a tree equipped with a total order
on its internal nodes.

In this paper, an $(m,n)$ {\em shuffle} on symbols $p$ and $q$ is simply a
sequence of length $m+n$ containing $m$ $p$'s and $n$ $q$'s.  (The
complete terminology for such a sequence is \emph{riffle shuffle}
\citep{shuffleAldousDiaconis86}.) For example $p q p p q$ is a $(3,2)$
shuffle on $p$ and $q$.  The utility of these shuffles in the present
context is summarized in the following observation.
\begin{obs}
  \label{obs:shuffle}
  Given totally-ordered sets $P$ and $Q$, the total
  orderings of $P \cup Q$ respecting the given orderings of $P$ and
  $Q$ are in one-to-one correspondence with the $(|P|,|Q|)$ shuffles
  on symbols $p$ and $q$.  
\end{obs}
To see how this works, assume total orders $p_1 < p_2 < \cdots <
p_m$ on $P$ and $q_1 < q_2 < \cdots < q_n$ on $Q$ are given, along
with a $(m,n)$ shuffle on $p$ and $q$. The required total ordering on
$P \cup Q$ is obtained by progressing along the shuffle and
substituting $p_i$ and $q_j$ for $p$ and $q$ in order: for example the shuffle
$p q p p q$ uniquely defines the total order $p_1 < q_1 < p_2 < p_3 < q_2$ when
$p_1 < p_2 < p_3$ and $q_1 < q_2$. In
the other direction, a total ordering on $P \cup Q$ uniquely defines a
$(|P|,|Q|)$ shuffle and a total ordering on each of $P$ and $Q$. 

We can use shuffles to develop a recursive formulation of 
ranked phylogenetic trees. 
Assume that $v$ is an internal node of a tree and that the tree $T_v$
containing the descendants of $v$ is composed of two daughter subtrees
$L_v$ and $R_v$. Assume $L_v$ and $R_v$ have $m$ and $n$ internal nodes, respectively. 
We define a ``shuffle at an internal node'' $v$ to be a $(m,n)$
shuffle on symbols $\ell$ and $r$. 
Assume $L_v$ and $R_v$ are ranked subtrees, i.e. there is a total ordering
on the internal nodes of each of $L_v$ and $R_v$. 
By Observation~\ref{obs:shuffle}, a total ordering on the internal
nodes of $T_v$ respecting the orderings on the internal nodes of $L_v$
and $R_v$ is equivalent to a shuffle at the internal node $v$. 
Therefore we can recursively reconstruct the rank function for any
ranked tree given a shuffle at each internal node. We define a 
\emph{tree shuffle} to be such a choice of shuffles. With Observation
\ref{obs:shuffle}, we have the following result, which is crucial to
our analysis:

\begin{obs}
  \label{obs:shufflerank}
Each rank function on a given tree being equally likely is
equivalent to the statement: For each internal node $v$, each shuffle
at $v$ is equally likely.
\end{obs}

\SBsection{Neutral models for ranked trees}
\label{sec:uniform}
In this section we formulate two classes of neutral models. First we
introduce the constant across lineages models, which are an obvious
generalization of the coalescent/Yule models. Then we introduce the
constant relative probability models, which generalize coalescent/Yule
models in a new direction. The common theme between these two classes
of models is that they both induce the uniform distribution on
shuffles. 
\SBsubsection{Ranked oriented trees}
In this section it will be convenient to discuss {\em ranked oriented
trees} rather than ranked phylogenetic trees.  This allows us to distinguish the children of each vertex without having to explicitly label species which may later become extinct.
The distributions and
statistics considered in this paper may be easily transferred between
these two types of trees, as well as ranked unlabeled trees, which we
call \emph{ranked tree shapes}.
For the present purposes, definitions and proof are much easier in the
case of ranked oriented trees.

\begin{defi}
A {\em oriented tree} is a finite rooted binary tree where
the children of each internal node are labeled {\em left} and {\em
right} respectively. A {\em ranked oriented tree} is a oriented tree with a rank function.
\end{defi}
Note that here ``tree'' in this definition is not short for
``phylogenetic tree.'' That is, our notion of ranked oriented tree
does not include leaf labels. 
These trees are called ``oriented'' because they are oriented graphs,
i.e. the edges around each vertex have a fixed orientation.

There is a map from ranked oriented trees to ranked tree shapes
which forgets the ordering of children.  Similarly, there is a map
from ranked phylogenetic trees (called {\em ranked trees} in this
context) to ranked tree shapes which forgets leaf labels.
Both of these ``forgetful'' maps induce maps between probability
distributions on these sets of trees.

We now prove that the uniform distributions on ranked oriented trees and on ranked phylogenetic trees induce the same distribution on ranked tree shapes.

Recall that at {\em cherry} is a pair of adjacent leaves: two leaves
with the same parent.

\begin{prop}
Given a ranked tree shape with $n$ leaves and $k$ cherries, there are $2^{n-1-k}$ ranked oriented trees sent to it by the forgetful map.
\end{prop}
\begin{proof}
First, note that the $n-1$ internal nodes of a given ranked tree shape
are distinguished by their rank.  Thus every ranked oriented tree which maps to this tree shape must be formed by assigning an orientation at each of these internal nodes.
For each of these there are two possible left-right labelings of the child subtrees, giving $2^{n-1}$
ranked oriented trees.  However, for the $k$ internal vertices which are the parents of a
pair of leaves the ordering of children does not effect the resulting
ranked oriented tree.  For all other $n-1-k$ internal vertices the ranking of
internal vertices ensures that the two orderings of child subtrees are
distinguishable.  Thus there are exactly $2^{n-1-k}$ distinct ranked oriented trees.
\end{proof}

\begin{prop}
Given a ranked tree shape with $n$ leaves and $k$ cherries, there are $\frac{n!}{2^k}$ ranked phylogentic trees sent to it by the forgetful map.
\end{prop}
\begin{proof}
Similarly to the previous proof, there are $n!$ ways to label the
identified leaves of a ranked tree shape.  However, the labels of the two
leaves of a cherry may be switched without changing the ranked
phylogentic tree.  Such switches (and their combinations) are the only
such transformations which leave the phylogenetic tree unchanged.
Thus, there are $\frac{n!}{2^k}$
distinct leaf labelings of the ranked tree shape.
\end{proof}

Together, these give us the desired result:
\begin{prop} \label{PropUnif}
A uniform distribution on ranked oriented trees with $n$ leaves and a
uniform distribution on ranked phylogenetic trees with $n$ leaves both
induce the same distribution on ranked tree shapes.
\end{prop}
\begin{proof}
Given a ranked tree shape with $n$ leaves and $k$ cherries, there
are $2^{n-1-k} = 2^{n-1}/2^k$ ranked oriented trees which map to this
tree and $n!/2^k$ ranked phylogenetic trees which map to this tree.
Thus, for both of the induced probabilities on ranked tree shapes,
the ratio between the probability of a tree with $k$ cherries and the
fixed tree with $1$ cherry is $1/2^{k-1}$.  As both distributions are
probabilities they must be equal.
\end{proof}

\begin{prop}\label{PropNumROTs}
There are $(n-1)!$ ranked oriented trees on $n$ leaves.
\end{prop}
\begin{proof}
Proceed by induction on $n$; for $n=2$ the statement is obviously
true. Suppose there are $(n-1)!$ ranked oriented trees (ROTs) with $n$
leaves.  Now note that the next bifurcation event is uniquely identified by its rank, and it can occur in $n$ places, thus there will be $n!$ ROTs with $n+1$ leaves. \end{proof}

It is well known that the Yule model gives the uniform distribution on
ranked phylogenetic trees on $n$ taxa for each $n$ \citep{Edwards1970}.  By
Proposition \ref{PropUnif}, this corresponds to the uniform distribution on ranked
oriented trees.  A direct proof goes as follows:

\begin{prop}
The Yule model results in the uniform distribution on ranked oriented
trees with $n$ leaves after $n-1$ bifurcations.
\end{prop}
\begin{proof}
Given a ranked oriented tree with $n$ leaves, there are $n$ possible
places for a bifurcation event to occur.  These are all equally likely.  By induction the $(n-1)!$ ranked oriented trees with $n$ leaves were all equally likely, so the $n!$ ranked oriented trees with $n+1$ leaves are also all equally likely.
\end{proof}

The following lemma will be useful shortly.
\begin{lem}
\label{lemma:uniform_extinction}
Given a ranked oriented tree (ROT) with $n$ leaves, there are $n(n+1)$
ways to add an additional leaf. 
\end{lem}
\begin{proof}
First, decide which rank the new internal node will have, from $1$
(earliest) to $n$ (latest).
If the new internal node has rank $k$ then there are $k$ choices at
that level for the edge to add it to, and then $2$ choices for which
side of this edge the new pendant leaf will sit.  This gives a total
of $2 \sum_{i=1}^{n} i = 2 \frac{n(n+1)}{2} = n(n+1)$ ways to insert
the new leaf edge.
\end{proof}

In the sequel, we consider  {\it forward time
ranked-oriented-tree-valued stochastic processes}.  In particular we
consider birth-death processes where the transitions between trees
involve either an bifurcation or deletion event (e.g. speciation or
extinction).  Call these {\em ranked-oriented-tree birth-death
processes}.  The details of bifurcation (birth) and extinction (death)
events are as follows.  If there is a bifurcation event, in which two
pendant leaves are attached to an existing leaf, the new branches
descending from the bifurcation event are assigned \emph{left} and
\emph{right}.  If there is an extinction event, occurring at a leaf
vertex, the leaf and its adjacent edge  are deleted. The ancestor of
the extinct leaf is now a degree-two vertex.  This vertex is
suppressed by replacing it and its two adjacent edges by a single
edge, with an orientation for the new edge inherited in the obvious
way. In this way, the ancestor still has a left and right child.
The ranking of internal vertices is induced by the time ordering of their
associated bifurcation events.  If extinction events do not occur then
call such a process a {\em pure-birth ranked-oriented-tree process}.

In the later section on example applications we consider, for
computational convenience, the likelihood of rank functions on tree
shapes rather than on oriented trees. The following proposition and
corollary show that the uniform distributions on rank functions of
oriented trees in the models and analysis which follow induce uniform
distributions on tree shapes when orientation is forgotten.  In
particular, $p$-values for such rank functions may be computed over either oriented or unoriented tree shapes.

First, define a {\em symmetric vertex} to be one for which the unoriented shapes of the subtree below each child of this vertex are the same (isomorphic as tree shapes). A {\em big symmetric vertex} is a symmetric vertex with more than two leaves below.

\begin{prop}
A uniform distribution on rank functions on a given oriented tree induces a uniform distribution on rank functions of its corresponding tree shape.
\end{prop}
\begin{proof}
Let $t$ be an oriented tree with $n$ leaves and $t'$ its corresponding tree shape.
Let $q$ denote the number of big symmetric vertices of $t$.
For $n=2$, which implies $q=0$, we have for the ranking on $t'$ exactly $1 = 2^0$ ranking on $t$.
Assume the following statement is true for all oriented trees with
less than $n$ leaves:
for each ranking on $t'$ there are exactly $2^q$ rankings on $t$
which are sent to it by the map which forgets orientation of vertices.
The induction now breaks into three cases.

Case 1: Suppose the two children of the root branch-point of  $t'$ are
non-isomorphic tree shapes, each having more than $1$ leaf.  They may therefore be distinguished from
each other, and given a ranking on $t'$ the shuffle at the root node
of $t$ is determined.  Call the two child subtrees ``left'' and ``right''
with $q_1$ and $q_2$ big symmetric vertices, respectively.
By the inductive assumption, there are $2^{q_1}$ rankings for the left
subtree of $t$ and $2^{q_2}$ rankings for the right subtree which map
to the corresponding rank function on the left and right subtree
shapes.  This gives $2^{q_1+q_2}$ total since there is no choice for
the shuffle at the root branch-point of $t$.  This is the number of big
symmetric vertices of $t'$.

Case 2: Suppose the two children of the root branch-point of  $t'$ are
non-isomorphic tree shapes, one of the children being a leaf. They may therefore be distinguished from
each other, and given a ranking on $t'$ the shuffle at the root node
of $t$ is determined.  The bigger subtree $t_b$ has $q_b$ big symmetry vertices. By the inductive assumption, there are $2^{q_b}$ rankings for $t_b$ which map
to the corresponding rank function on its
shape. Attaching a leaf to $t_b$ to obtain $t$ does not change the number of rankings for $t$ or $t'$. Therefore, there are $2^{q_b}$ rank functions on $t$ which map to the given
rank function on $t'$, and $q_b$ is the number of big symmetric
vertices of $t'$.

Case 3: Suppose that the two children of $t'$ are isomorphic.  Therefore they may not be distinguished except by the ranking.
Therefore the shuffle at the root branch-point of $t$ is only
determined up to swapping the left and right subtrees.  After this
choice the two subtrees are distinguished: which subtree of $t'$ is
``left'' and which is ``right'' is determined by the shuffle.
The rest of the argument proceeds as before, except that this time
there are $2^{q_1+q_2+1}$ rank functions on $t$ which map to the given
rank function on $t'$, and $q_1+q_2+1$ is the number of big symmetric
branch-point of $t'$.

The result now follows by induction.
\end{proof}

\begin{cor}
If a probability function on ranked oriented trees is uniform on rank functions conditioned on oriented tree then it is also uniform on rank functions of an (unoriented) tree shape when conditioned on that (unoriented) tree shape.
\end{cor}
\begin{proof}
This follows from the previous proposition because the resulting mixture of uniform distributions on rank functions on $t'$ (one for each oriented tree $t$ with shape $t'$) is also uniform.
\end{proof}

This corollary allows us to apply our rank tests to trees which are given without orientation-- ranked tree shapes.

\SBsubsection{Constant across lineages (CAL) models}
We define a {\em constant across lineages} (CAL) model to be a forward time
ranked-oriented-tree birth-death process such that
any new (bifurcation or extinction) event is equally likely to occur
in any extant lineage.  You may also think of the projection of this
process onto ranked trees, by forgetting the orientation of children at each
internal vertex.
Any model described in terms of rates is a CAL
model if the bifurcation and extinction rates are equal between
lineages at any given time. However, these rates may vary in an
arbitrary fashion depending on time or the current state of the
process. This class of models includes the Yule model \citep{Yule1924},
the critical branching process model \citep{AlPo2005}, the constant
rate birth and death process \citep{Nee1994} and the coalescent
\citep{Kingman1982}.

However, the CAL class is more general.  It includes macroevolutionary
models that have global speciation and extinction rate variation, for
example due to global environmental conditions. Furthermore, it is
also possible to incorporate models which take into account incomplete
random taxon sampling, which is equivalent to the deletion of $k$
species uniformly at random from the complete tree.  Indeed, if the
complete tree evolved under a CAL model then we simply run the model
for longer with the probability of bifurcation set to zero and the
extinction probability non-zero (and uniform across taxa).  This
extended model is clearly still within the CAL class.

The CAL class also includes microevolutionary models such as the
coalescent with arbitrary population size history. This very simple
but important fact means that the tests for non-neutral
diversification described in later sections are not fooled by
ancestral population size variation (as are a number of other tests in
the literature).

The CAL definition is a generalization of the ``exchangeable''
criterion from \citet{Aldous2001}, and we acknowledge the importance
of Aldous' ideas in formulating the definition.

\begin{prop} \label{PropCAL1}
At all times in a CAL model, the distribution of ranked oriented trees
with $n$ leaves is uniform.
\end{prop}
\begin{proof}


Assume that after $k$ events, all $(m-1)!$ ranked oriented trees
(ROTs) of size $m$ are equally likely.  If the next event is a
bifurcation then, because the result of each (tree, bifurcation event)
pair is distinct, after this event all $m!$ ROTs with $m+1$ leaves are
equally likely.  Similarly, if the next event is an extinction then
for each of the $(m-1)!$ equally likely trees there are $m$ equally
likely choices for which leaf to extinguish, giving $m!$ possibilities
in all.  By Lemma \ref{lemma:uniform_extinction} each ROT with $m-1$
leaves results from $m(m-1)$ of these tree-plus-leaf choices.  Thus
each ROT with $m-1$ leaves is equally likely, with probability $m(m-1)
* 1/m! = 1/(m-2)!$.

Since this is true for any such sequence of bifurcations and extinctions it is true at all times.
\end{proof}

Recall that this corresponds to a uniform distribution on ranked
phylogenetic trees, by Proposition \ref{PropUnif}.

Of course, any model giving the uniform distribution on ranked trees
with $n$ tips gives the uniform distribution on rank assignments given
a topology with $n$ tips. Thus 
\begin{cor}
  \label{cor:calUniform}
Any CAL model gives the uniform
distribution on rank assignments (and thus tree shuffles) given a tree
topology.
\end{cor}

We have the following limited converse of Proposition \ref{PropCAL1}.
\begin{prop}  \label{PropCAL2}
Pure-birth CAL models are the precisely the set of pure-birth ranked-oriented-tree processes which, for any $n\ge 1$, give the uniform distribution on ranked oriented trees with $n$ taxa when halted as soon as $n$ taxa are present.
\end{prop}

\begin{proof}
By the proof of Proposition \ref{PropCAL1}, pure-birth CAL models result
in a uniform distribution on ranked oriented trees of size $n$ (since there
have been exactly $n-1$ events).

Now consider a model which does not satisfy the CAL condition.
Assume that the $k$-th bifurcation event was not picked uniformly
among lineages, i.e. there is a ranked tree $T_0$ with lineages $l_1$ and $l_2$ which have probabilities $p_1 \neq p_2$ to speciate.
Let $T_1$ (respectively $T_2$) be the ranked tree produced if $l_1$ (respectively $l_2$) bifurcates.  In a pure birth process, $T_1$ and $T_2$ may only be reached in this way.
  Now
  \[
  \bP[T_1] = \bP[T_0] \cdot p_1 \neq \bP[T_0] \cdot p_2 = \bP[T_2]
  \]
  which shows that this model cannot give the uniform distribution on
  ranked trees when the process is halted at $k$ taxa.  There is only one way to build each ranked oriented tree with $n$ leaves so the distribution on these cannot be uniform, since an equal number must descend from each of $T_1$ and $T_2$. Thus, by contradiction, there is no such $k$ and so no such model.
\end{proof}

Note that in the last proposition, the restriction to a pure-birth
process is needed. Consider a process with extinction where
bifurcation is equally likely for each species but extinction is
history dependent: whenever an extinction event occurs, it
undoes the most recent bifurcation event.
This model clearly does not belong to the class of CAL models.
However, it gives a uniform distribution on ranked trees of some fixed size.

\SBsubsection{Constant relative probability (CRP) models}
The motivation for the {\em constant relative probability} (CRP)
models comes from
considering the models on ranked trees which might emerge from
non-selective diversification, perhaps based on physical or
reproductive barriers. For example, assume we could watch a set of
species emerge via allopatric speciation, and the fundamental
geographic barrier is a mountain range dividing land into two regions,
$A$ and $B$. These regions may differ in size or fecundity, so there may be some
difference in the rate of diversification in $A$ versus $B$. However, our
neutral assumption for the CRP class is that the \emph{relative} rate stays
constant over
time. In contrast, non-neutral models might dictate that a bifurcation
in one region will shift the equilibrium such that further
diversification in that region will become more likely (``bursting''
diversification) or less likely (``refractory'' diversification).

Again, for convenience, we work with ranked oriented trees so we may distinguish the two children of any bifurcation event.
For each internal node, $v$, (representing a bifurcation event) let
$L_v$ and $R_v$ denote the ``left'' and ``right'' lineages descending from $v$ (daughter
subtrees of $v$).


A {\em constant relative probability} ({\em CRP}) model is a
forward time pure-birth ranked-oriented-tree process together with a
probability distribution $P$ on the unit interval $[0,1]$, where
each internal vertex has a real number, $p_v$ associated with it.
Each new bifurcation occurring in the clade below $v$ occurs in $L_v$
with probability $p_v$, and occurs in $R_v$ with probability $1-p_v$.
For each new bifurcation event (internal vertex), $v$, choose the value
$p_v$ by an independent draw from $P$.
As with CAL models, there is no constraint of any kind on waiting
times between bifurcation events.

Recall the map from ranked oriented trees to unranked oriented trees
which forgets the rank ordering of internal nodes and the leaf labels.
The image of a ranked tree under this map is its oriented tree. Similarly, if the orientation is also forgotten then call the
image the tree shape of the initial tree.

\begin{prop}
A CRP model, stopped at a time depending only on the time and number
of leaves, gives the uniform distribution on rank functions for each
oriented tree.
\end{prop}

\begin{proof}
Consider the distribution of ranked oriented trees resulting from the stopped
CRP.  Consider a particular oriented tree, $t$, with $k$ internal vertices
$v_1,\ldots, v_{k}$. Let $n_i$ and $m_i$ denote the number of internal vertices below the left
and right subtrees, respectively, of vertex $v_i$.
Fix a ranking on this tree.
We now compute the probability of this ranked oriented tree under the model
(conditional on the total number of leaves).
Fix an assignment of $p_{v_i}$ to each internal vertex $v_i$.
Given this choice, the probability of the given ranked tree is the
product of the probabilities of
each bifurcation event.
For a bifurcation at vertex $v_i$, the probability of this event is
the product of $p_{v_j}$ for all $v_j$ for which $v_i$ lies on its
left subtree times the product of $(1-p_{v_j})$ for all $v_j$ for
which $v_i$ lies on its right subtree.  In the product of these
probabilities over all $v_i$, the term $p_{v_j}$ occurs exactly $n_j$
times (once for each internal vertex on the left subtree of $v_j$) and
the term $(1-p_{v_j})$ occurs exactly $m_j$ times (once for each
internal vertex on the right subtree of $v_j$).
Thus, the probability of this ranked tree (given the choice of $p_v$) is:
$$\prod_{j=1}^k p_{v_j}^{n_j} (1-p_{v_j})^{m_j}$$
Note that this is independent of the ranking.
Since the $p_{v_i}$ are picked independently from a distribution $P$,
the probability of this ranked tree shape is
$$\int_{p_{v_1}}\cdots \int_{p_{v_{k}}} \prod_{j=1}^k p_{v_j}^{n_j} (1-p_{v_j})^{m_j} dP \cdots dP$$
which is again independent of the ranking.
Therefore, all rankings of this oriented tree are equally likely.
\end{proof}

Note that the CRP generalizes the stick-breaking models
\citep{aldousCladograms95}.  Recall that with the stick-breaking model, a
stick is recursively broken into pieces, with the break point of each
piece chosen independently from a probability $P$ on the open unit
interval $(0,1)$.  For example, if the number chosen for a piece was
$1/2$ then that piece is broken into two equal sized pieces.  For each
piece a new draw is taken from $P$ to determine the how far along to
break it.
To generate a finite oriented binary tree with $n$ leaves, first break a
stick as just described then choose $n$ points from the unit interval
uniformly at random.  This determines a consistent set of partitions
corresponding to a binary tree.

It is well known that the Yule model is generated by setting $P$ to
the uniform measure on $(0,1)$.  Similarly, Aldous's Beta model
corresponds to setting $P$ to be a beta measure.

Thus, the CRP process produces oriented ranked versions of such trees in
a sequential growth process.

\SBsection{Tests for bursting diversification based on shuffles}
\label{sec:test}

In the previous section, we demonstrated that any model satisfying the
constant across lineages or constant relative probability criteria
induces the uniform distribution on tree shuffles.  In this section we
describe a way of testing for deviation from the uniform distribution
on tree shuffles, and thus test for deviation from these neutral
models. We emphasize that this can go significantly beyond testing the
coalescent/Yule model, which is typically considered to be
the definition of neutrality.  Indeed, rejection of the uniform
distribution on shuffles rejects all of the CAL and CRP models
simultaneously, and the coalescent/Yule model is only one model in
these classes. We note further that although the focus of this section is to
consider all of the shuffles of a ranked tree at once, one can also
consider a shuffle at a particular node as described in the
introduction.

There are several useful tools available to test whether a shuffle
is likely to have come from the uniform distribution on shuffles.  In
fact, a number of tests in the statistics literature 
have been developed for testing equality of distributions which
actually implement a test of deviation from the uniform distribution
for shuffles.  These tests work as follows: assume we are given two sets
of samples $\{\ell_i\}_{i=1,\ldots,m}$ and $\{r_j\}_{j=1,\ldots,n}$
and would like to test the hypothesis that they are draws from the
same distribution. To test, combine the draws and put the samples in
increasing order (assume that all draws are distinct). This clearly
gives a shuffle on symbols $\ell$ and $r$.  If the draws are from
identical distributions then the induced distribution on shuffles will be
uniform; if on the other hand symbols cluster together in the shuffle,
there is some evidence that the draws are from unequal distributions. 

One can then test deviation from the uniform
distribution on shuffles in one of several ways. One way is to count the number of
``runs.'' As described in the introduction, a run is simply a sequence within the shuffle using only one symbol;
the shuffle $\ell \ell r r r r \ell$\, has three runs. Let $X_{m,n}$ denote the number
of runs under the uniform distribution on shuffles on $m$ symbols of
one type and $n$ of another.
The distribution of $X_{m,n}$ is classical (see, e.g. \citet{HoggCraig}):
\begin{equation}
  \begin{split}
  \bP\{X_{m,n}=2k+1\} &= \frac{{m-1 \choose k}{n-1 \choose k-1}+ {m-1
  \choose k-1}{n-1 \choose k}}{{m+n \choose m}} \\
  \bP\{X_{m,n}=2k\} &= \frac{2{m-1 \choose k-1}{n-1 \choose k-1}}{{m+n \choose m}}.
\end{split}
\label{EqnRunDistr} 
\end{equation}
Asymptotic results for the mean and variance are also known:
$$\bE[X_{m,n}]=\mu_{m,n} = 2\frac{mn}{m+n}+1, \qquad {\rm Var}[X_{m,n}]=\frac{(\mu_{m,n}-1)(\mu_{m,n}-2)}{m+n-1}.$$

The usual application of the runs test makes a shuffle from the two
draws as described above, calculates the number of runs in the
shuffle, and then uses the above-calculated probabilities to 
test deviation from the uniform distribution on shuffles. However, the
same method can be applied in any situation to test deviation
from the uniform distribution on shuffles. In the present case, we can
use an analogous process to investigate tree shuffles.

As described in the introduction, a tree shuffle simply assigns a
shuffle of appropriate size to each internal node of the tree; from
the previous section we expect these shuffles to be distributed
uniformly for a variety of neutral models.  Using runs we can test
whether a single shuffle is drawn from the uniform distribution, but
some method is needed to combine this information across the internal
nodes of the tree.

We chose to combine our data from each vertex by simply summing
the number of runs across all of the shuffles in the corresponding tree shuffle.
Let $\run(T)$ denote this number.
The distribution of $\run(T)$ (under the assumption that each shuffle
is equally likely) can be calculated recursively as shown in the next
several paragraphs. There are two cases to consider. First, one may
condition on the observed tree
topology and calculate the neutral distribution of $\run(T)$ in that
setting. A second option is to test deviation from a neutral model
which gives the uniform
distribution on ranked trees. This is a stronger statement than
saying that a given model induces the uniform distribution on shuffles
conditioned on the phylogenetic tree.

We first condition on the observed tree. Uniform shuffles conditioned
on the tree shape are obtained in the CRP and the CAL class of models.
For a tree with one leaf, we have $\bP\{\run(T) = 0\} = 1$. For a tree
with two leaves, we also have $\bP\{\run(T) = 0\} = 1$ (the two
daughter subtrees have no internal nodes).

For a tree $T$ with uniform random ranking, composed of two ranked subtrees $L$
and $R$ of size $m$ and $n$, respectively, we have:
\begin{equation} \label{eqn:TreeRuns}
  \bP\{\run(T) = k\} = \sum_{i=0}^{k} \bP\{X_{m,n} = i\}
  \sum_{j=0}^{k-i} \
  \bP\{\run(L) = j\} \bP\{\run(R) = k-i-j\}.
\end{equation}
It is shown in the Appendix that this distribution can be calculated
on a tree with $n$ leaves in time $O(n^3 \log^2 n)$.
Thus it is practical to obtain a $p$-value for $\run(T)$ analytically.

Now we take the second approach, assuming we want to test a model such
that each ranked tree is equally likely. This includes the CAL models,
and in the case of pure birth models, this is exactly the set of the CAL
models  (Proposition~\ref{PropCAL2}).
Let $\run(n)$ be the random variable ``runs of a tree with $n$ leaves'' where the tree is drawn from the uniform distribution on ranked trees.
The distribution of $\run(n)$ can again be obtained recursively. Note
that for a uniform ranked tree on $n$ leaves, the probability that one
daughter tree has size $r$ and the other daughter tree has size $n-r$
is $1/(n-1)$ for all $r$. Thus
\begin{equation}
  \begin{split}
  \bP\{\run(n) = k\} = \frac{1}{n-1} \sum_{r=1}^{n-1} \sum_{i=1}^k
  & \bP\{X_{r,n-r} = i\} \times \\
  & \sum_{j=0}^{k-i} \
  \bP\{\run(r) = j\} \bP\{\run(n-r) = k-i-j\}.
\end{split}
  \label{eq:TreeRunsYule}
\end{equation}
The complexity for recursively calculating the distribution of runs for trees with 
$n$ leaves is $O(n^4 \log^2 n)$,  by an argument analogous to that for 
Equation (\ref{eqn:TreeRuns}).

Note that there are a number of alternative ways to ``sums of runs'' for testing
deviation from the uniform distribution on shuffles. First, we have
made one choice--- namely, summation--- concerning how the statistics
for each shuffle are combined. One certainly could use an
alternative method, potentially including weights. Second, there are
other statistics such as Mann-Whitney-Wilcoxon which could be used in
place of the runs statistic. The advantage of summation is that
it results in simple formulas, and the advantage of the runs statistic is
that it is easy to interpret. We have not tested any alternate
formulations.

A \texttt{python} package for computing quantiles of shuffles is available at
\begin{verbatim}
http://www-m9.ma.tum.de/twiki/bin/view/Allgemeines/TanjaGernhard
\end{verbatim}

One of the main features of this package is the ability to calculate
the quantile of the runs statistic assuming a uniform distribution on
rankings for a set of input trees. The quantiles
can be calculated conditioned on a given tree shape, or under the
assumption of a uniform distribution on ranked trees.  For a
collection of trees (e.g. a sample from the Bayesian posterior), the
individual quantiles can be averaged.  In addition to the calculation
of the runs statistic and the quantile for the whole tree, the package
can calculate the runs statistic and quantile for each interior vertex
of a single tree.  This feature may be useful for biologists looking
for signals of a key innovation.


\SBsubsection{Shuffles in the Bayesian setting}

In our work up to now, we have assumed that the correct tree and
diversification timing information is known. This assumption is not
realistic for a number of datasets. For example, below we apply our
methodology to a sample of Hepatitis C viruses, which probably do not
have enough sequence divergence to perfectly reconstruct a
phylogenetic tree with timing information. 

One way of working with such datasets is to take a Bayesian approach,
where rather than a single tree one gets a posterior distribution on
trees. For each single tree, one can compute the quantile of the total
runs statistic, either conditioning on the topology or assuming a Yule
distribution of ranked tree shapes. We then simply take the average of
the quantiles thus computed for each tree. Such averaging can be
justified in a manner similar to the work of
\citet{drummondSuchard07}, except that no further simulation is needed
to compute the $p$-value. The average of $p$-values in this case is
not exactly uniformly distributed under the neutral model as a proper
$p$-value should be, although the averaged distribution does share
many of the characteristics of a classical $p$-value \citep{mengPosteriorPredictive94}.

\SBsubsection{Runs and neutrality}

Here we note that the runs statistic can be used to test
the coalescent in the presence of ancestral population size variation.  
Tests of neutrality in the presence of historical population
size variation are of particular recent importance, as new
coalescent-based methods are in use to infer population size history
in a Bayesian framework
\citep{bayesianSkylineDrummondEA05,opgenRheinEA05}. If these methods
are to be used on a given set of sequences it is important to test the
central assumption of the methods, namely that the sequences have a
genealogy which can be accurately described using the coalescent with
arbitrary population size history. 

 Unfortunately, classical statistics such as the
$D$ statistics of \citet{dTajima89} and \citet{dFuLi93} confound
ancestral population size changes and non-neutral evolution. One
solution to this problem is to investigate the Bayesian posterior on
phylogenetic trees for evidence of non-neutral evolution rather than
using the sequence information directly. This has been done by
\citet{drummondSuchard07}, who use a posterior predictive $p$-value
approach. Here we simply point out that, as described above, the
coalescent with arbitrary population size history is a CAL model and
thus will induce the uniform distribution on ranked phylogenetic
trees; thus by rejecting the CAL class we reject a general coalescent
model. We will apply this fact below in the example application to
Hepatitis C data.

\SBsubsection{Generalization for non-binary trees}
 
Polytomies (i.e. non-binary splits) are common in reconstructed
phylogenetic trees. Some polytomies are certainly due to a lack of
information to resolve the splits, however it has been argued that
molecular and species level polytomies actually exist
\citep{jackmanLizards99,polytomiesSlowinski01}. 
The methodology described in this paper can be extended to trees
with ``hard'' polytomies, i.e. cases of multiple divergence which are
essentially simultaneous in evolutionary time.

The new ingredient needed is the ``multiple runs distribution,'' i.e.
the analog of (\ref{EqnRunDistr}) for
shuffles on more than two symbols. This is described in
\citet{MR0155371}. Using these distributions, the probability of a
shuffle consisting of symbols from the $k$ daughter trees
can be found for a shuffle at a non-bifurcating split $v$.

\SBsection{Example applications}

In this section we describe two distinct applications of the methods in this
paper. First, we apply the methods to E1 gene data for
the Hepatitis C virus (HCV). This data set shows some limited-- though
consistent-- lineage-specific bursting diversification, showing that
neither CAL nor CRP models accurately describe the sort of evolution
observed. However, an
analysis not conditioning on tree shape clearly rejects any CAL model,
such as the coalescent with varying population size. The second
application is to phylogenetic trees for ants, whose timing information
was reconstructed through fossils and the r8s \citep{sandersonR8s03}
rates smoothing program. These ant trees do not show any evidence of
lineage-specific bursting evolution, despite some
interesting history in terms of diversification rates.

Our HCV data comes from two independent studies: one in
China \citep{luEaHCVChina05}, and one in Egypt \citep{rayHCV00}. 
The HCV alignments were
retrieved from the LANL HCV database \citep{lanlHCV} via PubMed
article ID
numbers. The Chinese dataset contained samples from 132 infected
individuals, and the Egyptian dataset had samples from 71 individuals.
We randomly partitioned the taxa from the Chinese dataset into three
sets of 44 taxa each and used the corresponding sub-alignments as
distinct data sets. The Egyptian data was similarly split into two
sub-alignments of size 37 and 36. This partitioning was done in order
to have a larger number of similar datasets from which we could investigate
the dynamics of HCV evolution, and to demonstrate that non-neutral
evolution can be seen even with an moderate number of taxa.

In order to
avoid confounding temporal information with molecular rate variation, we
applied the relaxed clock model of \citet{relaxedPhyloDrummondEA06}
as implemented in the BEASTv1.4 suite of computer programs \citep{beast}.
We chose uncorrelated lognormally distributed local clocks, the HKY
model, and four categories of gamma rate parameters in the gamma +
invariant sites model of sequence evolution. We used both the constant
population size and exponential growth coalescent priors. All
other parameters were left as default; the corresponding BEAST XML
input files are available from the authors upon request.

In each case the MCMC chain was run for 10 million generations, and
convergence to stationarity checked with the BEAST program Tracer. For
each model parameter, the minimum effective sample size (ESS) was at
least 164, with most being significantly greater. The coefficient of
variation of the relaxed clocks in the analysis had a minimum of 0.336
and an average of 0.491, indicating a significant deviation from a
strict clock for this data set. The first 10\% of the run was removed
and 100 trees were taken from the tree log file, equally spaced along
the run of the MCMC chain. We interpret these trees as being
independent samples from the posterior. As a check, the analysis was
run with an empty alignment and no consistent deviation from the
uniform distribution on shuffles was detected (results not shown).

\eatfig{
\begin{table}
  \centering
  \begin{tabular}{c||cc|cc}
    Data set & Const. cond. & Exp. cond. & Const. CAL & Exp. CAL \\
    \hline
    China set 1 & 0.232 & 0.254 & 0.0415 & 0.0601 \\
    China set 2 & 0.191 & 0.17 & 0.041 & 0.0349 \\
    China set 3 & 0.239 & 0.259 & 0.0287 & 0.0265 \\
    Egypt set 1 & 0.261 & 0.299 & 0.045 & 0.0624 \\
    Egypt set 2 & 0.308 & 0.256 & 0.0242 & 0.0188 \\
  \end{tabular}
  \caption{Expected quantiles of the number of runs in the posterior
  for a Bayesian analysis as described in the text. Each row
  represents one dataset. ``Const.'' means the
  BEAST analysis with a constant population coalescent prior, and
  ``Exp.'' denotes analysis an exponentially increasing
  population size coalescent prior. The ``cond.'' label means that we analyze
  conditional on tree topology, which gives us thus the quantile for any
  neutral model inducing the uniform distribution on shuffles. ``CAL''
  denotes the runs quantile under the assumption of the uniform distribution on ranked trees,
  as would be the case for any CAL model, such as the coalescent with
  arbitrary population size history. As described in the text, the
  ``cond.'' columns show that some limited lineage-specific bursting is
  seen, and the CAL column rejects the coalescent with arbitrary
  population size history.}
  \label{tab:e1hcv}
\end{table}
}

We have displayed the results in Table~1. 
In the columns labeled ``cond.'' we show the quantile of the number of runs conditioning on tree
shape, calculated as in Equation (\ref{eqn:TreeRuns}).
As can be
seen, the results are substantially below one half, with the maximum
being 0.308. Although this is not exceptionally
strong lineage-specific bursting behavior, it does so consistently
across five samples from two independent studies. Thus we feel
confident in saying that the evolution of HCV displays
lineage-specific bursting behavior. It might also be noted that these
results were gained despite the fact that the coalescent was used as a
prior. That is, if any bias could be expected in the Bayesian
analysis, it would be towards a coalescent prior and a uniform
distribution on shuffles, thus we believe our results form an
upper bound for the actual statistics of the HCV lineages.

\eatfig{
\begin{figure}
  \begin{center}
    \includegraphics[height=4.5cm]{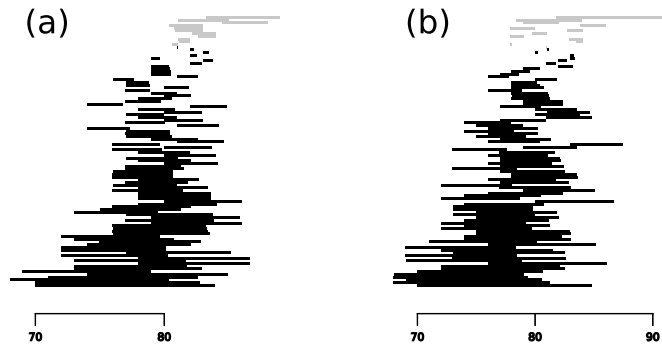}
  \end{center}
  \caption{A visualization of the number of runs in a posterior sample
  of trees for an alignment of Hepatitis C sequence data of
  \citet{rayHCV00}.  Black bars represent fewer runs than neutral, and
  gray the opposite.  As described in the text, the width of the bars
  represents the amount of divergence from a broad class of neutral
  models.  Said simply, each
  black bar represents a tree in the posterior which displays evidence of
  lineage-specific bursting diversification, and the longer the bar, the more bursting
  the tree. The bars are sorted vertically by increasing size (with
  sign.) Figure (a) shows the results when the tree
  prior in BEAST was taken to be coalescent with constant population size. Figure (b)
  shows the corresponding results with the exponentially increasing
  population size prior.} 
  \label{fig:hcv}
\end{figure}
}

We have displayed a graphical representation of the results for the
second Chinese data set in Figure~3. Each
horizontal bar
represents one of the 100 ranked trees from the posterior. One side of the bar
gives the number of runs in the ranked tree $T$, and the other side gives the
expected number of runs for a neutral (i.e. CAL or CRP) tree of the same unranked topology as
$T$. If $T$ has more runs than the expectation, the bar is colored
gray; if fewer it is colored black. In both the cases of constant
population size and exponentially increasing population size
coalescent prior for BEAST, it can
be seen that there are fewer runs than the expectation, meaning that
it appears that the HCV data under investigation may have had periodic
bursts of diversification in its past. 

Now we apply our techniques as a statistical test for the coalescent
with ancestral population size variation as described above. 
This is topical: we note that the \citet{rayHCV00} HCV data was analyzed by \citet{opgenRheinEA05} as an example
application of a reversible-jump Bayesian MCMC algorithm for
estimating demographic history of the virus. In doing so they made an
implicit assumption of neutrality because their method [and other
such methods \citep{bayesianSkylineDrummondEA05}] are based on the
coalescent. They did not test this neutrality assumption as no methods
were available at the time to test for neutral evolution in the
presence of ancestral population size changes. 

Our method can do so. Specifically, we compare the number of runs to
the distribution for an arbitrary CAL model, as in Equation (\ref{eq:TreeRunsYule}).
By the results in the right half of
Table~1, one can see that the data does not 
follow a coalescent model with arbitrary population size history. 
This implies a significant model mis-specification in the
\citet{opgenRheinEA05} paper; it would be interesting to know how this 
would impact the historical population size estimates in their paper.

For the second application we investigated two different trees of ant
taxa. The first tree is that of \citet{moreauEaAnts06}, showing the
diversification of the major ant lineages. The timing information in
this tree is quite remarkable, in that the corresponding
lineages-through-time (LTT) plot shows a substantial increase in
diversification rate during the Late Cretaceous to Early Eocene, which
corresponds to the rise of angiosperms (flowering plants). Given the tools at our
disposal, one might wonder if this increase in diversification rates
affected all lineages equally, or if it occurred in lineage-specific
bursts.  The second ant tree we investigated was that of \Pheidole, a
``hyperdiverse'' ant genus. Pheidole is almost certainly monophyletic,
and yet comprises about 9.5\% of the ant species in the world,
according to latest estimates \citep{moreauPheidole}.
Moreau has recently reconstructed a phylogeny of this
genus which we have analyzed along with the tree of the ant lineages
in general. Both trees were reconstructed via maximum likelihood, then
made ultrametric using the penalized likelihood method of the r8s
rates smoothing program \citep{sandersonR8s03}. 

\eatfig{
\begin{figure}[h]
  \begin{center}
    \includegraphics[height=5cm]{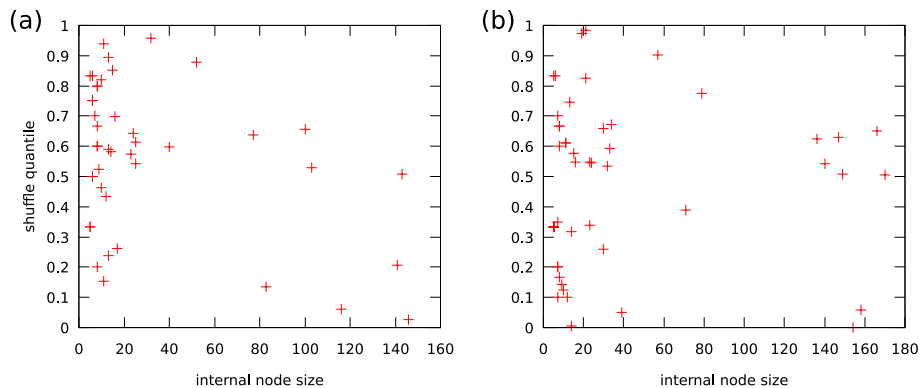}
  \end{center}
  \caption{The distribution of the runs statistic
  for the internal nodes in two trees of ant taxa.  Each point in
  each plot represents an internal node in the corresponding tree; the
  $x$ axis gives the number of taxa below the internal node and the $y$ axis gives the quantile of that
  internal node in terms of the runs statistic. Figure~(a) is the tree
  of \citet{moreauEaAnts06}, and Figure~(b) is a tree of Pheidole.
  These two trees do not appear to consistently show either
  lineage-specific bursting or refractory diversification.}
  \label{fig:bothAnts}
\end{figure}
}

In Figure~4 we show a plot of the internal nodes of
each tree. The $x$ coordinate in the plot is the
number of taxa below an internal node, and the $y$ axis is the quantile of the number
of runs in the shuffle statistic. As can be seen, there is no clear
correlation between number of taxa below an internal node and the shuffle statistic, and
at no stage does diversification appear to be consistently bursting or
refractory in a lineage-specific sense. We can also compute the
quantile of the total number of runs across the tree: for tree (a)
this is about 0.9052 and for tree (b) this is about 0.6718.  Thus for
these two ant trees we do not see any significant evidence of
lineage-specific bursting or refractory diversification. This analysis
forms an interesting counterpoint to the LTT results for the ants,
which shows an overall increase in diversification in rate during the
Late Cretaceous to Early Eocene across the entire tree.

\SBsection{Conclusion}

We have developed a framework which allows testing for non-neutral
diversification timing. Our work consists of three main components:
first, a simple, recursive way of quantifying the relative timing information
on a phylogenetic tree; second, two classes of neutral models on trees
with timing information; and third, a summary statistic which allows
comparison of reconstructed trees to these neutral models.  In our
methodology, timing information is considered relative to sister taxa
and considered in the context of the tree, which may make it a valuable
complimentary method to lineages-through-time plots.
We compute the significance of the deviation of timing information
from a neutral model analytically, using a simple method drawn from
classical statistics.

This method was conceived for the macroevolutionary case, in order to
find historical evolutionary patterns requiring explanation. However,
it is also quite applicable in the microevolutionary case,
where it can test neutrality in the presence of historical population
size variation. This is particularly relevant as methods are becoming
available to describe historical population size under a coalescent
assumption.

We emphasize that our methodology can go substantially beyond testing
for deviation from the constant rate birth and death models,
which are usually the entire class of ``neutral'' models considered.
Indeed, because \emph{any} CAL or CRP model induces the uniform
distribution on shuffles, deviation from this distribution is evidence
to reject any model in the CAL or CRP classes. Such a conclusion is much
stronger than deviation from a constant rate birth and death model, 
which is only one of the CAL models. 

However, sometimes one may wish to test only a more restricted set of
models, such as only the CAL models (which include the coalescent with
arbitrary population size history) and not the more general CRP models.
By testing a more restricted class of models, a particular dataset
will be more likely to fall outside the chosen class. For example, in
the application of our methods to Hepatitis C data above, the data
consistently shows evidence of not coming from a CRP model, although
the corresponding quantile is in the 0.17 to 0.31 range. However, if
one tests for conformity to the CAL class (again, including the
coalescent with arbitrary population size history) one obtains
rejection at the 5\% level.

We applied the methods on two types of data. First we investigated
data for the Hepatitis C virus, demonstrating that it consistently shows evidence
of some limited lineage-specific bursting evolution. We then applied
our methods to reject the coalescent with arbitrary population size
history for this data. Next, we investigated two phylogenetic trees
of ants, which showed no evidence of unusual relative diversification
rates, despite an interesting history of overall diversification
rates.

We recall that our method uses
``relative'' timing information rather than actual branch lengths. 
In many ways this is an advantage. 
In a microevolutionary setting this means that the corresponding tests
are invariant to changes in ancestral population size, and thus our
test for neutrality is not ``fooled'' by ancestral population size
variation.
In a macroevolutionary setting the statistics are robust to branch length
estimation error over long time scales. Such estimations are known to
be difficult \citep{kimuraEvolDist81}.
We note further that from a modeling perspective
it is possible to specify a probability distribution on ranked
phylogenetic trees without specifying a particular distribution on
branch lengths. This flexibility means that it may be possible to
reject many models at once as described above. 

Nevertheless it may be useful at some future stage to combine topology and
continuous branch length information, rather than the discretized
version considered here.  However, quantifying the shape of such
objects appears to be challenging, as the relevant geometry is quite
intricate \citep{billeraHolmesVogtmann01,orangesMoultonSteel04}.  In
contrast, by discretization to ranked trees we obtain a purely
combinatorial object. 

We close by noting that although various techniques for reconstructing
phylogenetic trees with timing information have been present for many
years, these methods are currently seeing an intense period of
development and will only improve. With this improvement we expect to
see an increase in the number of trees present in the literature with
interesting patterns of diversification timing due to adaptive
radiation or other factors. We hope that our technique will
prove to be a useful analytical tool for these future investigations,
not only for finding interesting diversification patterns, but also
for testing potential biases of timing reconstruction methods.

\SBsubsubsection{Acknowledgements}

The idea of using ranked trees in this context was suggested by John Wakeley. 
The authors would like to thank Alexei Drummond, Arne Mooers, Allen
Rodrigo, and Dennis Wong for helpful discussions. Mike Steel made
several important suggestions, including the recursive calculation of
$p$-values for the shuffles. Alexei Drummond gave very valuable
advice on running BEAST. John Huelsenbeck generously allowed us to run
BEAST on his cluster. Corrie S. Moreau supplied trees for analysis,
and gave many very helpful comments on the manuscript.  TG is funded by a PhD
scholarship of the Deutsche Forschungsgemeinschaft.  FAM is funded by
the Miller Institute for Basic Research in Science at the University
of California, Berkeley. 

\SBsection{Appendix}

Here we provide a proof of the time-complexity bound for the
computation of the runs distribution $\run(T)$ (i.e. conditioning on a
given tree shape).  This distribution may be computed easily for certain tree shapes,
such as the comb tree.  However, here we provide a bound which holds for all tree shapes.
This bound makes use of a bound on the number of runs in a ranked tree.

Let $r(n)$ denote the maximum number of runs for a ranked tree with $n$ leaves.
Thus $r(1) = r(2) = 0$, $r(3) = 1$ and $r(4)=2$.
Let $I_{i=n/2}$ be $1$ if $i=n/2$ and $0$ otherwise.
For a tree with at least $2$ leaves, if the first branch point has $i$ leaves on one side and $n-i$
leaves on the other, with $i \le n-i$, then the number of runs at this vertex may be up to
$2(i-1) + 1 - I_{i=n/2}$ (note that we have an $(i-1,n-i-1)$ shuffle at this vertex).  This maximum is obtained by a shuffle which interleaves the elements from
each set, one from each side for as long as possible, starting with the largest side.

Thus, $r(n)$ satisfies the following recurrence:
$r(1)=r(2)=0$ and for $n\ge 2$:
\begin{equation}
r(n) = \max_{1\le i \le n/2} \left( 2i - 1   - I_{i=n/2} + r(i) + r(n-i) \right) \notag
\end{equation}

\begin{prop}
For all integers $n\ge 1$, $r(n) \le n \log_2 n$.
\end{prop}
\begin{proof}
The statement is true for $n=1$.
Suppose that the statement is true for all $k<n$. Then,
\begin{eqnarray}
r(n) & = & \max_{1\le i \le n/2} \left( 2i - 1   - I_{i=n/2} + r(i) + r(n-i) \right) \notag \\
& \le & \max_{1\le i \le n/2} \left( 2i - 1 + i \log_2 i +  (n-i)\log_2 (n-i) \right). \notag
\end{eqnarray}
Note that $2i-1$, $i \log_2 i$ and $(n-i)\log_2 (n-i)$ are all convex
functions of $i$ so their sum is convex also.
Thus, the maximum of  $2i - 1 + i \log_2 i +  (n-i)\log_2 (n-i)$ occurs at an extreme value.
Setting $i=1$ gives $1 + 0 + (n-1)\log_2(n-1)$, while setting $i=\frac{n}{2}$ gives $2\frac{n}{2}-1 + 2\frac{n}{2} \log_2 \frac{n}{2} = n(\log_2 2 + \log_2 \frac{n}{2}) -1 = n \log_2 n -1$.  Both of these values are less than $n \log_2 n$ and so $r(n)$ must be at most $n \log_2 n$.
The result follows for all $n \ge 1$ by induction.
\end{proof}

We now proceed to bound the complexity of computing the distribution of runs for a tree.  For a tree $T$ with $1$ or $2$ leaves, the number of runs is always $0$.

Let $T$ be a tree with $n \ge 3$ leaves; we assume a uniform
distribution on tree shuffles.
Let $L$ and $R$ be the two randomly ranked subtrees of $T$, with $a$ and $b$ leaves respectively.

Equation (\ref{eqn:TreeRuns}) may be rewritten as follows:
\begin{eqnarray} \label{eqn:TreeRunsNarrow}
  \bP\{\run(T) = k\} & = & \sum_{i=0}^{A_1} \bP\{X_{a,b} = i\} \sum_{j=0}^{A_2} \
  \bP\{\run(L) = j\} \bP\{\run(R) = k-i-j\} \notag \\
  & = &\sum_{i=1}^{A_1} \bP\{X_{a,b} = i\} \bP\{\run(L) + \run(R) = k - i\} \label{eqn:TreeRunsNarrow2}
\end{eqnarray}
where $A_1 = \min(k, n)$ and $A_2 = \min(k-i, r(a))$. Note that $a+b=n\ge 3$ implies $X_{a,b} \ge 1$ and $\run(T) \ge 1$.

Since $\run(T)$ is supported on (i.e. zero outside of) $k = 1,\ldots, \lfloor n \log_2 n \rfloor$, and  for each $k$, the computation of Equation (\ref{eqn:TreeRunsNarrow2}) costs $2n-1$ operations,
the cost of computing its distribution with this formula is at most $(\lfloor n \log_2 n \rfloor)(2n-1)$ 
arithmetic operations plus the cost of computing
$ \bP\{X_{a,b} = i\}$ for $i=1,\ldots, n$ and $\bP\{\run(L) + \run(R) = x\}$ for $x = 0, \ldots, r(n) - 1 \le n \log_2 n - 1$.

For these fixed $a$ and $b$, the values of $\bP\{X_{a,b} = i\}$ can be calculated 
using Equation (\ref{EqnRunDistr}) in constant time (at most $5*2+4 = 14$ arithmetic operations each) with a linear overhead as follows. 
The binomial coefficients ${a \choose k}$ for $a \leq b$ and $k \leq
b$ in Equation (\ref{EqnRunDistr}) may be calculated with at most two arithmetic operations from the factorials, $j!$ for $1\le j \le n$, which may in turn be pre-calculated in linear time ($n-1$ multiplications).
Thus, calculating $\bP\{X_{a,b} = i\}$ for $i=1, \ldots, n$ takes at most $14n$ arithmetic operations, with a one-time overhead of $n-1$.

The distribution of $\bP\{\run(L) + \run(R) = x\}$ is supported on $x=0, \ldots, \lfloor n \log_2 n \rfloor - 1$.
It may be computed by repeated application of the formula
\begin{equation}
\bP\{\run(L) + \run(R) = x\} = \sum_{j=0}^{\lfloor (n-1) \log_2 (n - 1) \rfloor} \bP\{\run(L) = j\} \bP\{\run(R) = x-j\} \notag
\end{equation}
as long as the distributions of $\run(L)$ and $\run(R)$ are know.
This computation requires at most 
$n \log_2 n \left( 2 (n-1) \log_2 (n-1) + 1\right)$
arithmetic operations: at most $(n-1)\log_2(n-1) +1$ multiplications and $(n-1)\log_2(n-1)$ additions for each of $n \log_2 n$ values of $x$.
Note that the distribution of $\bP\{\run(L)\}$ is supported by $j=0, \ldots, {\lfloor (n-1) \log_2 (n - 1) \rfloor}$, since $L$ has at most $n-1$ leaves.

So, if the distribution of $\run(L)$ and $\run(R)$ are known, the distribution of $\run(T)$ may be calculated in at most 
$$(\lfloor n \log_2 n \rfloor )(2n-1) +  14n + n \log_2 n \left(2(n-1) \log_2 (n-1) +1\right)$$
 arithmetic operations.  This is at most
 $$2n^2 \log_2 n + 2 n^2 \log_2^2 n + 14n $$
 for all $n\ge 3$.  Since $\run(T)$ is $0$ for $n=1,2$ the time to calculate it is $0$.

This procedure may be applied recursively, computing the distribution
of runs of all subtrees before finally computing the run distribution
of $T$.  Since there are $n-1$ internal vertices and each has at most
$n$ leaves below it, the total number of arithmetic operations
required is at most $n(2n^2 \log_2 n + 2 n^2 \log_2^2 n + 14n+1)$
(including the overhead for pre-computing $j!$).  This is $O(n^3
\log^2_2 n)$.

\end{spacing}

\newpage
\bibliographystyle{sysbio}
\bibliography{ranked}

\end{document}